\documentclass{aastex}
\usepackage{emulateapj5}
\usepackage{apjfonts,pstricks,amsmath,epsfig,pst-node}

\makeatletter
\newenvironment{inlinetable}{%
\def\@captype{table}%
\noindent\begin{minipage}{0.999\linewidth}\begin{center}}
{\end{center}\end{minipage}\smallskip}
 
\newenvironment{inlinefigure}{%
\def\@captype{inlinefigure}%
\noindent\begin{minipage}{0.999\linewidth}\begin{center}}
{\end{center}\end{minipage}\smallskip}
\makeatother 

\newcommand\Acon{A_{\rm contamination}}

\slugcomment{Submitted to {\it{The Astrophysical Journal}}}
\shortauthors{Gonzalez et al.}
\shorttitle{The LCDCS Correlation Function}
 
\begin{document}
\title{The Las Campanas Distant Cluster Survey - The Correlation Function}
  
\author{Anthony H. Gonzalez\altaffilmark{1}}
\affil{ Harvard-Smithsonian Center for Astrophysics, 60 Garden Street,
   Cambridge, MA 02138}
\altaffiltext{1}{Present Address: Department of Astronomy, University of
Florida, P.O. Box 112055, Gainesville, FL 32611}   

\author{Dennis Zaritsky}
\affil{Steward Observatory, University of Arizona, 933
   North Cherry Avenue, Tuscon, AZ 85721}
 
\author{Risa H. Wechsler\altaffilmark{2}}
\affil{Department of Physics, University
   of California, Santa Cruz, CA 95064}
\altaffiltext{2}{Present Address: Physics Department, University of Michigan,
Ann Arbor, MI 48109. }       

\begin{abstract}                                                   

We present the first non-local ($z>$0.2) measurement of the
cluster-cluster spatial correlation length, using data from the Las
Campanas Distant Cluster Survey (LCDCS).  We measure the angular
correlation function for velocity-dispersion limited subsamples of the
catalog at estimated redshifts of $0.35\le z_{est}<0.575$, and derive
spatial correlation lengths for these clusters via the cosmological
Limber equation.  The correlation lengths that we measure for
clusters in the LCDCS are consistent both with local results for the
APM cluster catalog and with theoretical expectations based upon the
Virgo Consortium Hubble Volume simulations and the analytic
predictions.  Despite samples containing over 100 clusters, our ability 
to discriminate between cosmological models
is limited because of statistical uncertainty.

\end{abstract}

\section{Introduction}

The spatial correlation function of galaxy clusters provides an
important cosmological test, as both the amplitude of the correlation
function and its dependence upon mean intercluster separation are
determined by the underlying cosmological model. In hierarchical
models of structure formation, the spatial correlation length, $r_0$,
is predicted to be an increasing function of cluster mass, with the
exact dependence
determined by $\sigma_8$
(or equivalently $\Omega_0$, using the constraint on
$\sigma_8-\Omega_0$ from the local cluster mass function) and the
power spectrum shape parameter, $\Gamma$.  Low density and low $\Gamma$ models
generally predict stronger clustering for a given mass and a greater
dependence of the correlation length upon cluster mass.

The three-space correlation function of clusters was first measured for
subsamples of the Abell catalog by \citet{bah83} and \citet{kly83}.
Both groups found that the correlation function is well-described by a power
law, $\xi(r)=(r/r_0)^{-\gamma}$, and obtained
a correlation length
$r_0\simeq 25 h^{-1}$ Mpc with $\gamma$$\simeq$2.  \citet{bah83} also
observed a strong dependence of correlation strength upon cluster richness,
which was later quantified by \citet{bah88} and \citet{bah92w} as a roughly
linear dependence of $r_0$ upon $d_c$, the mean intercluster separation. 
\citet{pos92} and \citet{pea92} confirmed the form of the correlation function
for the Abell catalog in their larger spectroscopic samples, with 
both studies obtaining $r_0$$\simeq$20$h^{-1}$ Mpc for clusters with richness
class R$\ge$1.

While these correlation lengths have strong implications for cosmological 
models, a key problem with interpretation of the Abell results 
is concern
that the observed correlation lengths are positively skewed by 
projection effects and sample inhomogeneities
\citep[see][]{sut88,dek89,efs92,pea92}.
Several analyses find that $\xi(\sigma,\pi)$ is strongly anisotropic,
evidence that these effects are significant \citep{sut88,efs92,pea92}. 
Still, the net impact of these factors is unclear. Contrary to the
concerns raised by these studies, 
\citet{mil99} use an expanded
sample of Abell clusters to derive correlation lengths that are
consistent with earlier analyses and robust to projection effects, and
\citet{van97} argue
that projections are insufficient to account for the stronger correlation 
observed in the Abell catalog as compared to the APM catalog \citep{dal92}.

Fortunately, independent 
constraints on the correlation function have arisen as new catalogs with
automated, uniform selection criteria have become available
\citep[e.g.,][]{dal92,nic92,dal94,nic94,rom94,cro97,collins2000,mos2000}. 
Some of the recent optical catalogs, such as the APM \citep{dal92}, also
probe to lower $d_c$ than the Abell samples, while the X-ray samples provide
greatly improved leverage for the most massive (highest $d_c$) clusters.
While
systematic variations persist between samples, the existing data are generally 
consistent with $r_0$ slowly increasing with $d_c$ 
\citep[however, see][]{collins2000}.  

To exploit the growth of observational data, there has been a corresponding 
theoretical effort to predict the cluster spatial correlation function as
a function of mass and epoch. Driven by 
cosmological volume N-body
simulations \citep[e.g.,][]{gov99,col2000,mos2000} and the development of a
well-tested analytic formalism \citep{mo96,she99,smt1999}, 
a theoretical framework has been established that enables derivation
of quantitative cosmological constraints from the observational data.
The mass dependence can be studied using existing data sets and is typically
best matched by
low-density models \citep[see][]{cro97,bor99b,collins2000}, although systematic uncertainties
and observational scatter have precluded precision cosmological constraints.
In contrast, the redshift dependence remains unconstrained because the data have not
existed to test the evolutionary predictions of these models.

In this paper we utilize the Las Campanas Distant Cluster Survey (LCDCS) to
determine the spatial correlation length at $z$$\simeq$0.45.
We first measure the angular correlation function for a series of
subsamples at this epoch and then derive the corresponding $r_0$ values via
the cosmological Limber inversion \citep{peebles80,efs91,hud96}. The resulting
$r_0$ values 
constitute the first 
measurement at this epoch of the dependence of the cluster correlation
length upon $d_c$, probing mean separations similar to previous local optical
catalogs.
Popular structure formation models predict only a small
amount of evolution from z=0.45 to the present, as illustrated in section 5. We
test this prediction by comparing our results with the local observations.  

\section{The Las Campanas Distant Cluster Survey}
\label{sec-lcdcs}

The recently completed Las Campanas Distant Cluster Survey, which
contains 1073 candidates, is the largest published catalog of galaxy
clusters at $z\ga0.3$ \citep{gon2001}.  Clusters are detected in the
LCDCS as regions of excess surface brightness relative to the mean sky
level, a technique that permits wide-area coverage with a minimal
investment of telescope time.  The final statistical catalog covers an
effective area of 69 square degrees within a $78\degr\times1.6\degr$
strip of the southern sky ($860\times24.5$ $h^{-1}$ Mpc at $z$=0.5 for
$\Omega_0$=0.3 $\Lambda$CDM). \citet{gon2001} also provide estimated
redshifts, $z_{est}$, based upon the brightest cluster galaxy (BCG)
magnitude-redshift relation that are accurate to $\sim$15\% at
$z_{est}=0.5$, and demonstrate the existence of a correlation between
the peak surface brightness, $\Sigma$, and velocity dispersion,
$\sigma$. Together these two properties enable construction of
well-defined subsamples that can be compared directly with simulations
and observations of the local universe.

\section{The LCDCS Angular Correlation Function}
\label{sec-angular}
We wish to measure the two-point angular correlation function 
for both the full LCDCS catalog and for well-defined
subsamples at $z$$\simeq$0.45 that we shall use to constrain the mass dependence of
the correlation length.  \citet{gon2001} find that the velocity 
dispersion is related to the peak surface brightness of the cluster
detection via
$\log\sigma\propto\log \Sigma(1+z)^{5.1}$, and so we employ this
relation to define subsamples that are roughly velocity dispersion
limited (see Figure \ref{fig:sbthreshold}).
Error in the redshift dependence of this relation can induce a systematic
bias in construction of our subsamples; however, this uncertainty is 
sufficiently small as to have negligible impact for the LCDCS data within
the redshift range of the subsamples. We restrict all subsamples to $z$$>$0.35
to avoid sample incompleteness at lower redshift, while the maximum
redshift ($z$=0.575) is set by sample size. Subsamples with larger 
redshift limits (and hence higher surface brightness limits) result
in smaller samples ($N$$<$100) and correspondingly larger statistical 
uncertainties.

To compute the two-point angular correlation function, we use the
estimator of \citet{lan93},
\begin{equation}
\omega(\theta)=\frac{n_r(n_r-1)}{n(n-1)}\frac{DD}{RR}-\frac{n_r-1}{n}\frac{DR}{RR}+1,
\label{eqn:lscor}
\end{equation}
which \citet{ker2000} have demonstrated is more robust than other
algorithms. In Equation \ref{eqn:lscor}, $n$ is the number of
clusters, $n_r$ is the number of random points, DD is the number of
cluster pairs with angular separation $\theta\pm\delta\theta/2$, RR is
the number of random pairs with the same separation, and DR is the
number of cluster-random pairs.  To construct 
\begin{inlinefigure}
\plotone{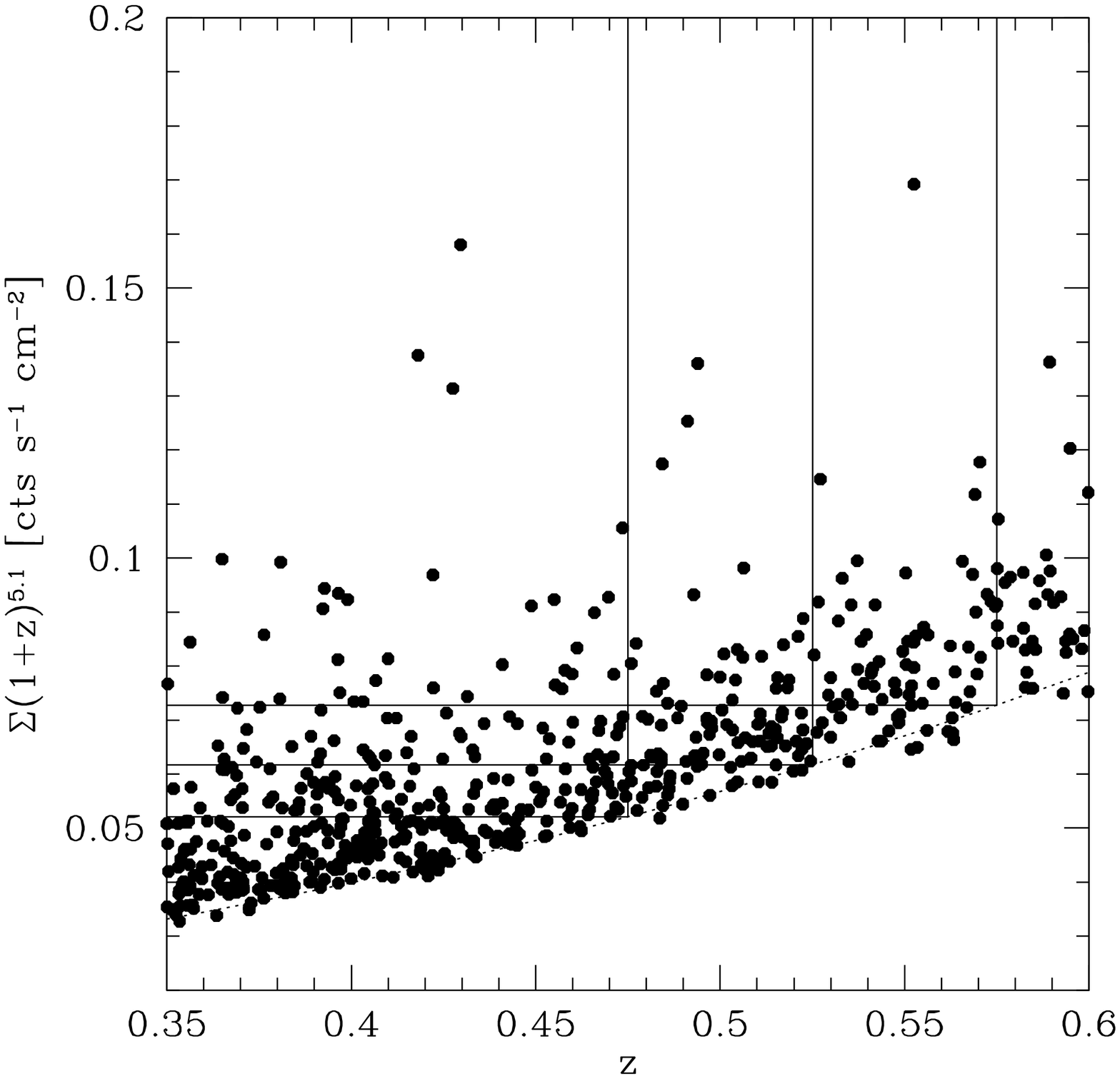} \figcaption[sbthreshold.eps]{Extinction-corrected
surface brightness distribution of the LCDCS clusters used to
construct subsamples. The solid lines denote the redshift and
$\Sigma(1+z)^{5.1}$ thresholds of the subsamples. The dotted curve is
the surface brightness threshold of the full LCDCS catalog for
E($\bv$)=0.05.  Candidates at the detection threshold in lower
extinction regions will lie slightly below this curve. Errors in the
estimated redshifts move individual data points along tracks similar
to this curve.
\label{fig:sbthreshold}}
\end{inlinefigure}            
\vskip 0.1cm
\noindent the random sample, we
first generate a list of 44,000 random, unmasked locations within the
survey region (2000 locations in each of the 22 scan sets that
comprise the LCDCS survey).  Next, we reject locations that fail any
of the automated cluster selection criteria described in
\citet{gon2001}, yielding a final random sample of 33,886 positions.

Modelling the correlation function as a power law,
\begin{equation}
\omega(\theta)=A_\omega \theta^{1-\gamma}=
\left(\frac{\theta}{\theta_0}\right)^{1-\gamma},
\end{equation}
we use a maximum likelihood approach to determine the best-fit values
for $A_\omega$ and $\gamma$. Similar to \citet{cro97} and \citet{bor99b}, we 
maximize the likelihood function for an assumed Poisson probability distribution,
\begin{equation}
{\cal{L}}=\prod_{i}^{N} p_i =\prod_{i}^{N} \frac{\mu_i^{n_i}
\exp(-\mu_i)}{n_i!},
\end{equation}
where $n_i$ and $\mu_i\equiv<DD>$  are the observed and expected number 
of pairs 
in the interval d$\theta$. The final results from the maximum likelihood 
analysis are not dependent upon the exact choice of d$\theta$, which can be
made almost arbitrarily small.
To determine the best-fit model parameters we use pairs with separations
of 2$\arcmin$ -- 5$\degr$, and Monte Carlo simulations are utilized to determine
the associated 1-$\sigma$ uncertainties.
Specifically, we use the best-fit parameters as initial values, generate
1000 random realizations of the cluster-cluster pair distribution, and then
use the distribution of recovered parameter values to quantify the
observational uncertainties.
The angular correlation function for the entire LCDCS catalog and for our
lowest redshift subsample are shown in Figure \ref{fig:angcf}, overlaid
with best-fit power law models. For the full LCDCS catalog,
we obtain $\gamma$=$-1.78\pm0.10$ and 
$A_\omega$=$-1.44\pm0.04$ ($\theta_0$=51$\pm$22$\arcsec$).

\begin{table*}
\begin{center}
\caption{Angular Correlation Lengths}\label{tab:cf}
\begin{tabular}{clllllll}
\hline\hline
\multicolumn{1}{c}{$z$} &
\multicolumn{1}{c}{$<$$z$$>$} &
\multicolumn{1}{c}{$N$} &
\multicolumn{1}{c}{$\log A_\omega$} &
\multicolumn{1}{c}{$\gamma$} &
\multicolumn{1}{c}{$f$} &
\multicolumn{1}{c}{$\log A_{\omega,cor}$} \\
\multicolumn{1}{c}{(1)} &
\multicolumn{1}{c}{(2)} &
\multicolumn{1}{c}{(3)} &
\multicolumn{1}{c}{(4)} &
\multicolumn{1}{c}{(5)} &
\multicolumn{1}{c}{(6)} &
\multicolumn{1}{c}{(7)}  \\
\hline
All        & 0.56 &1073 & $-1.44\pm$0.04 & 1.78$\pm$0.10 &  0.29 & $-1.18\pm$0.04 \\ 
0.35-0.475 & 0.42 & 178 & $-1.26\pm$0.17 & 2.15$\pm$0.19 &  0.14 & $-1.13\pm$0.19 \\ 
0.35-0.525 & 0.46 & 158 & $-1.35\pm$0.33 & 2.30$\pm$0.33 &  0.16 & $-1.20\pm$0.33 \\ 
0.35-0.575 & 0.50 & 115 & $-1.42\pm$0.47 & 2.51$\pm$0.45 &  0.19 & $-1.24\pm$0.45 \\ 
\\
All        & 0.56 &1073 & $-1.73\pm$0.05 & 2.15          &  0.29 & $-1.47\pm$0.05 \\ 
0.35-0.475 & 0.42 & 178 & $-1.26\pm$0.13 & 2.15          &  0.14 & $-1.13\pm$0.13 \\ 
0.35-0.525 & 0.46 & 158 & $-1.25\pm$0.14 & 2.15          &  0.16 & $-1.10\pm$0.14 \\ 
0.35-0.575 & 0.50 & 115 & $-1.14\pm$0.18 & 2.15          &  0.19 & $-0.96\pm$0.18 \\ 
\hline
\end{tabular}

\medskip
\begin{minipage}{0.60\linewidth}
{\small
Note --- (1) Redshift range used to measure the angular correlation. (2) Number of
clusters. (3) Values
of $\log A_\omega$ are for $\theta$ in degrees. 
(7) Contamination-corrected amplitude of the angular
correlation function.
The second set of parameters in the table is for fixed values of the slope $\gamma$.
}
\end{minipage}
\end{center}
\end{table*}

Table \ref{tab:cf} lists information for the full LCDCS catalog and
the three subsamples at $z$$\simeq$0.45, including the redshift range
spanned by each subsample, the mean redshift of the subsample, the best-fit values of $A_\omega$ and
$\gamma$, and the estimated fractional contamination.  We also 
present best-fit values for $A_\omega$ fixing 
$\gamma$=2.15 --- equivalent to the best-fit value for the lowest
redshift subsample and similar to the best fit value for the ROSAT 
All-Sky Survey 1 Bright Sample \citep[][$\gamma$=2.11$^{+0.53}_{-0.56}$]{mos2000}.

Because the LCDCS candidates are not spectroscopically confirmed,
we must correct the correlation amplitude $A_\omega$ for the impact
of contamination before this data can be used to derive the spatial
correlation length $r_0$. 
If we assume that the contamination is spatially correlated
and can be described by a power law with the same slope as the cluster
angular correlation function (a reasonable approximation because for
galaxies, which are likely the primary contaminant,
$\gamma$$\simeq$1.8-1.9 \citep[e.g.][]{roc99,cab2000}), then the
observed angular correlation function is
\begin{equation}
\omega(\theta) = A_\omega \theta^{1-\gamma} =
\left(A_{\rm cluster} (1-f)^2+\Acon f^2\right) \theta^{1-\gamma},
\end{equation}
where $f$ is the fractional contamination. For detections induced by
isolated galaxies of the same magnitude as BCG's at $z\simeq0.35$ (and
identified as galaxies by the automated identification criteria
described in \citealt{gon2001}), we measure that $A_{\rm gal}$ is
comparable to $A_{\omega_A}$, the net clustering amplitude for all
LCDCS candidates at 0.3$<$$z$$<$0.8. For detections identified as low
surface brightness galaxies (including some nearby dwarf galaxies) we
measure $A_{\rm LSB}\simeq10A_{\omega_A}$. While these systems are
strongly clustered, we expect that they comprise less than half of the
contamination in the LCDCS.  For multiple sources of contamination the
effective clustering amplitude $\Acon=\sum A_i f_i^2/(\sum f_i)^2$, so
the effective clustering strength of the contamination is $\Acon\la
2.5 A_{\omega_A}$ even including the LSB's.

The last column in Table \ref{tab:cf} gives the
contamination-corrected value of $A_\omega$, under the assumption that
$\Acon=A_{\omega_A}$.  If the effective correlation amplitude of the
contamination is in the range $\Acon=(0-2.5) A_{\omega_A}$ (which
corresponds to uncorrelated contamination for the lower limit), then
the systematic uncertainties in $A_\omega$ and the corresponding uncertainties
in the $r_0$ values derived
in \S\ref{subsec-cfres} are $\la1$\%. These uncertainties, which are
small because the contamination only contributes to the observed correlation
function with weight $f^2$, are far less than
the statistical uncertainties in all cases.  Uncertainty in the
fractional contamination of the catalog yields a larger 6\% systematic
uncertainty in $r_0$, which is also less than the typical statistical
uncertainty.
\begin{table*}[t]
\begin{center}
\caption{Spatial Correlation Lengths}\label{tab:rodc}
\begin{tabular}{lllllll}
\hline\hline
\multicolumn{1}{c}{$z$} &   
\multicolumn{2}{c}{$\Lambda$CDM ($\Omega_0=0.3$) }  & 
\multicolumn{2}{c}{OCDM ($\Omega_0=0.3$)} & 
\multicolumn{2}{c}{$\tau$CDM($\Gamma$=0.2)}   \\
\multicolumn{1}{c}{range} & 
\multicolumn{1}{c}{$d_c$} & \multicolumn{1}{c}{$r_0$} & 
\multicolumn{1}{c}{$d_c$} & \multicolumn{1}{c}{$r_0$} & 
\multicolumn{1}{c}{$d_c$} & \multicolumn{1}{c}{$r_0$}  \\
\hline
0.35-0.475  & 38.4 & 14.7$^{+2.0}_{-2.2}$ & 33.8 & 12.9$^{+1.7}_{-2.0}$ & 30.9 & 11.8$^{+1.5}_{-1.8}$ \\ 
0.35-0.525  & 46.3 & 16.2$^{+2.7}_{-3.3}$ & 40.6 & 14.2$^{+2.3}_{-2.8}$ & 36.9 & 12.9$^{+2.1}_{-2.5}$ \\ 
0.35-0.575  & 58.1 & 18.2$^{+4.0}_{-4.9}$ & 50.8 & 15.8$^{+3.4}_{-3.9}$ & 46.0 & 14.4$^{+3.1}_{-3.8}$ \\
\\ \hline
\end{tabular}
\medskip
\begin{minipage}{0.55\linewidth}
{\small
Note --- The units for $d_c$ and $r_0$ are $h^{-1}$ Mpc. Listed uncertainties are statistical.
}
\end{minipage}
\end{center}
\end{table*}

\section{The Spatial Correlation Length}
\label{subsec-limber}

The observed angular correlation function can be used to determine the
three-space correlation length if the redshift distribution of the
sample is known. This is accomplished via the cosmological Limber
inversion \citep{peebles80,efs91,hud96}. For a power-law correlation
function with redshift dependence $f(z)$,
\begin{equation}
        \xi(r)=\left(\frac{r}{r_0}\right)^{-\gamma}\times f(z).
\end{equation}
The corresponding comoving spatial correlation length is $r_0(z)=r_0
f(z)^{1/\gamma}$, and the Limber equation is
\begin{eqnarray}
r_o^\gamma & = & A_\omega\frac{c}{H_0}\frac{\Gamma(\gamma/2)}{\Gamma(1/2)\Gamma[(\gamma-1)/2]} \times \\
           &   & \left[
           \frac{\int_{z1}^{z2} (dN/dz)^2 E(z) D_{A}(z)^{1-\gamma} 
f(z) (1+z) dz}
                {\left(\int_{z1}^{z2} (dN/dz) dz\right)^2}\right]^{-1},
\end{eqnarray}
where $dN/dz$ is the redshift distribution of objects in the sample,
$D_{A}(z,\Omega_0, \Omega_\Lambda)$ is the angular diameter distance,
and
\begin{equation}
E(z)=[\Omega(1+z)^{3} + \Omega_R(1+z)^{-2} +\Omega_\Lambda]^{1/2}, 
\end{equation}
as defined in \citet{peebles93}.  Because little evolution in the
clustering is expected over the redshift intervals spanned by our
subsamples (see the Appendix and Figure \ref{fig:cfvslocal}), $f(z)$
can safely be pulled out of the integral.

We estimate  the true redshift distribution of clusters in each LCDCS
subsample, $dN/dz$,
based upon the observed distribution of estimated redshifts, $dN_{obs}/dz$.
Because the redshift distribution of the full LCDCS catalog is slowly varying
over the redshift interval probed by our subsamples,
we are able to derive approximate models for the $dN/dz$ of the
subsamples by \\
\begin{inlinefigure}
\plotone{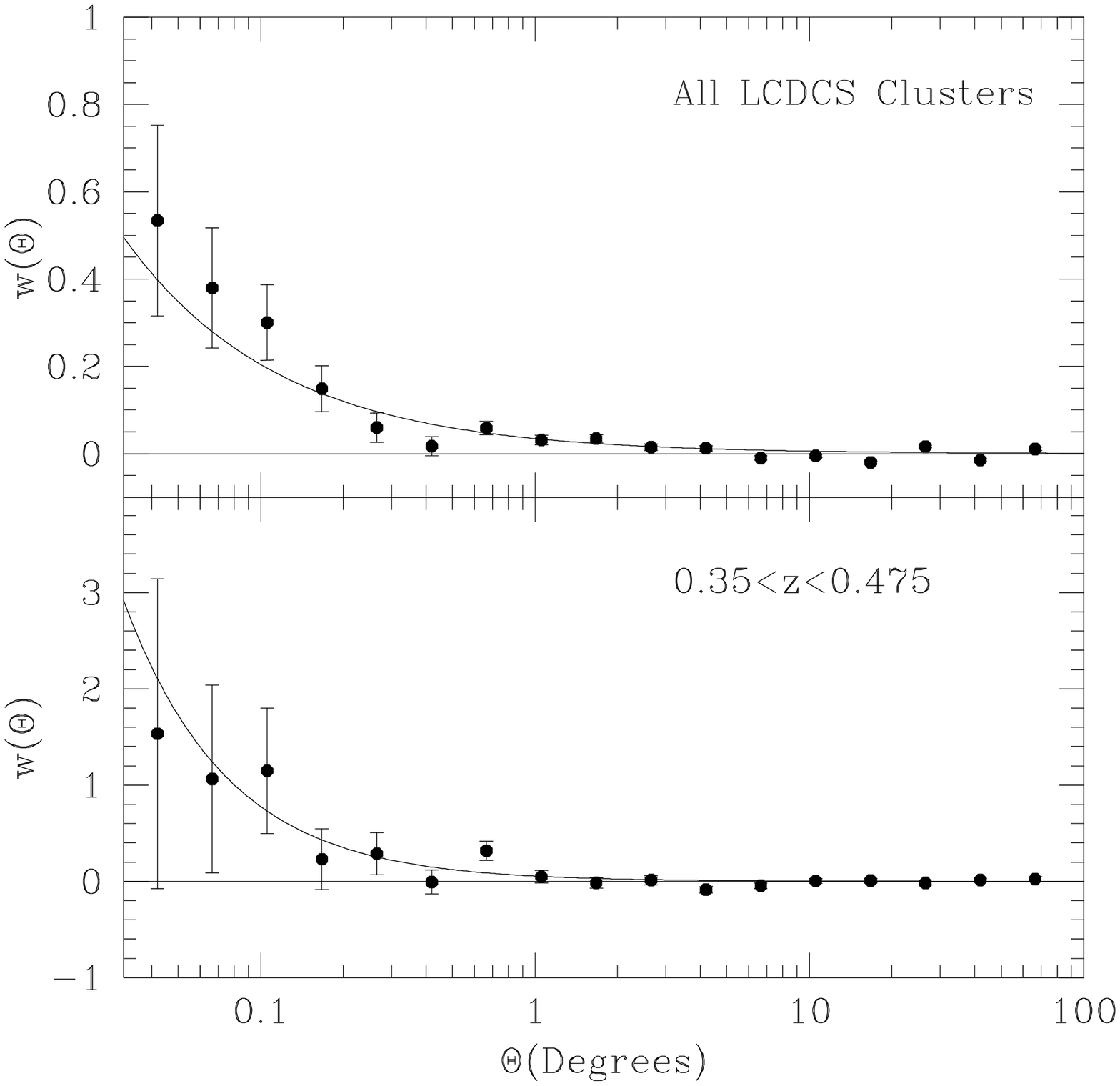} \figcaption[angcf.eps]{ Angular correlation functions
for the entire LCDCS catalog and for a volume-limited subsample with
$0.35\le z<0.475$.  The solid lines are best-fit power law models,
with the best-fit parameters given in Table \ref{tab:cf}. The data shown
are plotted with bins of width $\Delta\log \theta$=0.2; for the actual 
maximum likelihood analysis of the
subsamples we use $\Delta\log \theta$=0.005.
\label{fig:angcf}}
\end{inlinefigure}            

\vskip 0.35cm
\noindent convolving the $dN_{obs}/dz$ with Gaussian scatter 
\citep[$\sigma_z/z$$\approx$0.14 at $z$=0.5;][]{gon2001}, which eliminates
the sharp redshift limits imposed on $z_{est}$ in construction of the
subsamples.
The rigorously correct method for obtaining the $dN/dz$ is to 
{\it de}convolve the scatter; however, our approximation is sufficiently
accurate for use in the Limber inversion because of both the slow
variation of the LCDCS redshift distribution at this epoch
and the weak dependence of the Limber inversion upon $dN/dz$. 
To test the validity
of this approach, we also try modeling the $dN/dz$ using the theoretical
mass function of \citet{she99} convolved with redshift uncertainty.
Comparing these two methods we find that the derived spatial
correlation lengths agree to better than 3\% for all
subsamples. This result is insensitive to the exact mass
threshold assumed for the theoretical mass function, again due to the weak 
dependence of the Limber inversion upon $dN/dz$.

\section{Results and Comparison with Local Data}
\label{subsec-cfres}

Table \ref{tab:rodc} lists the correlation lengths ($r_0$) and mean
separations ($d_c$) that we derive for each of the three subsamples,
\begin{inlinefigure}
\plotone{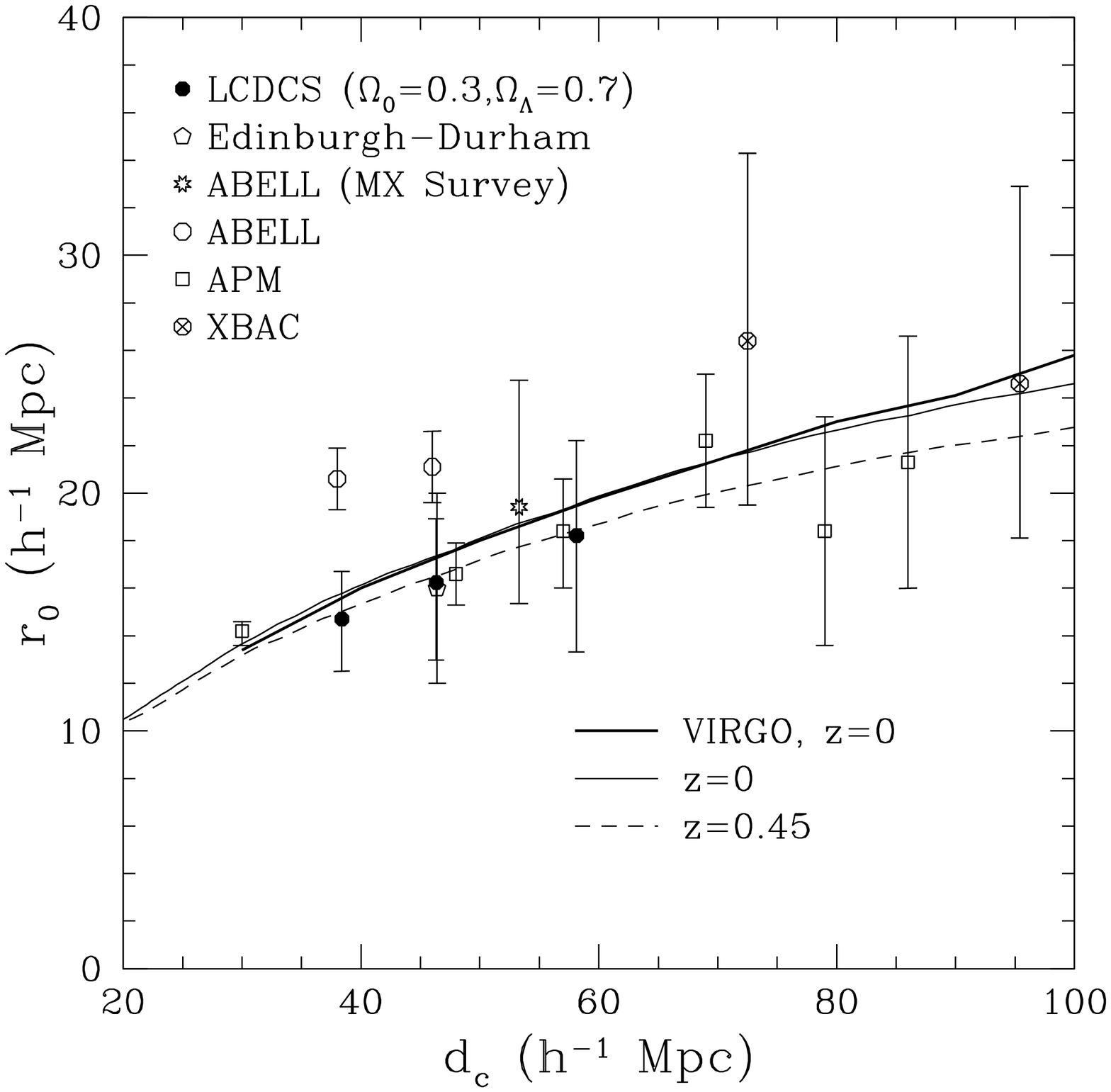}
\figcaption[fiducial] {Comparison of the LCDCS data with
local samples for a $\Lambda$CDM cosmology with $\Omega_0$=0.3.
The error bars on the LCDCS data correspond to the 1-$\sigma$ 
statistical uncertainty. Overlaid are the results from the Virgo
Consortium Hubble Volume simulations \citep{col2000} and analytic
predictions ($\Gamma$=0.2) for $z$=0 and $z$=0.45. These fiducial
curves do not account for the impact of uncertainty in cluster
masses (see Figure \ref{fig:apmcomp}).
\label{fig:cfvslocal}}
\end{inlinefigure}
\\
treating both $\gamma$ and $A_\omega$ as free parameters.
Both the $r_0$ and $d_c$ values are
cosmology-dependent, and so we list the derived values for
$\Lambda$CDM ($\Omega_0$=0.3), OCDM ($\Omega_0$=0.3), and $\tau$CDM
cosmologies. 
Values of $d_c$ in Table \ref{tab:rodc} are computed as $d_c\equiv
n^{-1/3}=[N(1-f)/V]^{-1/3}$, where $N$ is the number of clusters, $f$
is the contamination rate, and V is the effective comoving volume for
a given subsample. Including the impact of Gaussian scatter upon the
redshift distribution, the effective comoving volume for a subsample
with redshift bounds $z_1$ and $z_2$ is
\begin{equation}
V=\delta\Omega\times\int_{z_1}^{z_2}{dz\;p\frac{dV}{d\Omega dz}}
\end{equation}
The effect of the redshift uncertainty is contained in $p$, which is
the probability that a cluster at redshift $z$ will be observed to \\
\noindent have an estimated redshift within the redshift range spanned by the
subsample.  For Gaussian redshift scatter,
\begin{equation}
p(z)=\frac{\int_{z_1}^{z_2}{dz'\;e^{-(z'-z)^2/2\sigma_z^2}}}{
\int_{0}^{\infty}{dz'\;e^{-(z'-z)^2/2\sigma_z^2}}}.
\end{equation}
Inclusion of $p$ increases the computed values of $d_c$ by $\sim$3\%.

We compare the LCDCS results with several local studies to assess the degree of
evolution between $z$=0.5 and the present epoch. We find that the 
LCDCS $r_0$ values are
statistically consistent with results from
the Edinburgh-Durham Galaxy Catalogue
\citep{nic92}, APM survey \citep{cro97}, and the MX Survey
\citep[northern sample,][]{mil99}, but smaller than the correlation lengths
 found by
\citet{pea92} for their subsample of the Abell catalog. 
The correlation lengths from these studies are plotted in Figure
\ref{fig:cfvslocal}, where the LCDCS data points are shown for an assumed
$\Lambda$CDM cosmology ($\Omega_0$=0.3). 
The lowest $d_c$ data points for the XBAC catalog from \citet{aba98}, which probe
higher masses than our study, are also plotted. We also include fiducial
theoretical curves to illustrate the predicted degree of evolution 
between these epochs. The thick solid curve shows the $z$=0 results from
the Virgo Consortium Hubble Volume simulations ($\Omega_0 = 0.3$,
$\Omega_\Lambda= 0.7$, $\sigma_8=0.9$,$\Gamma$=0.17). The other
two curves correspond to analytic predictions for the dependence of $r_0$
on $d_c$ (or equivalently, on number density) at the two epochs,  using the analytic model
of \citet{she99} with $\Gamma$=0.2.

\begin{figure*}
\epsscale{0.65} 
\plotone{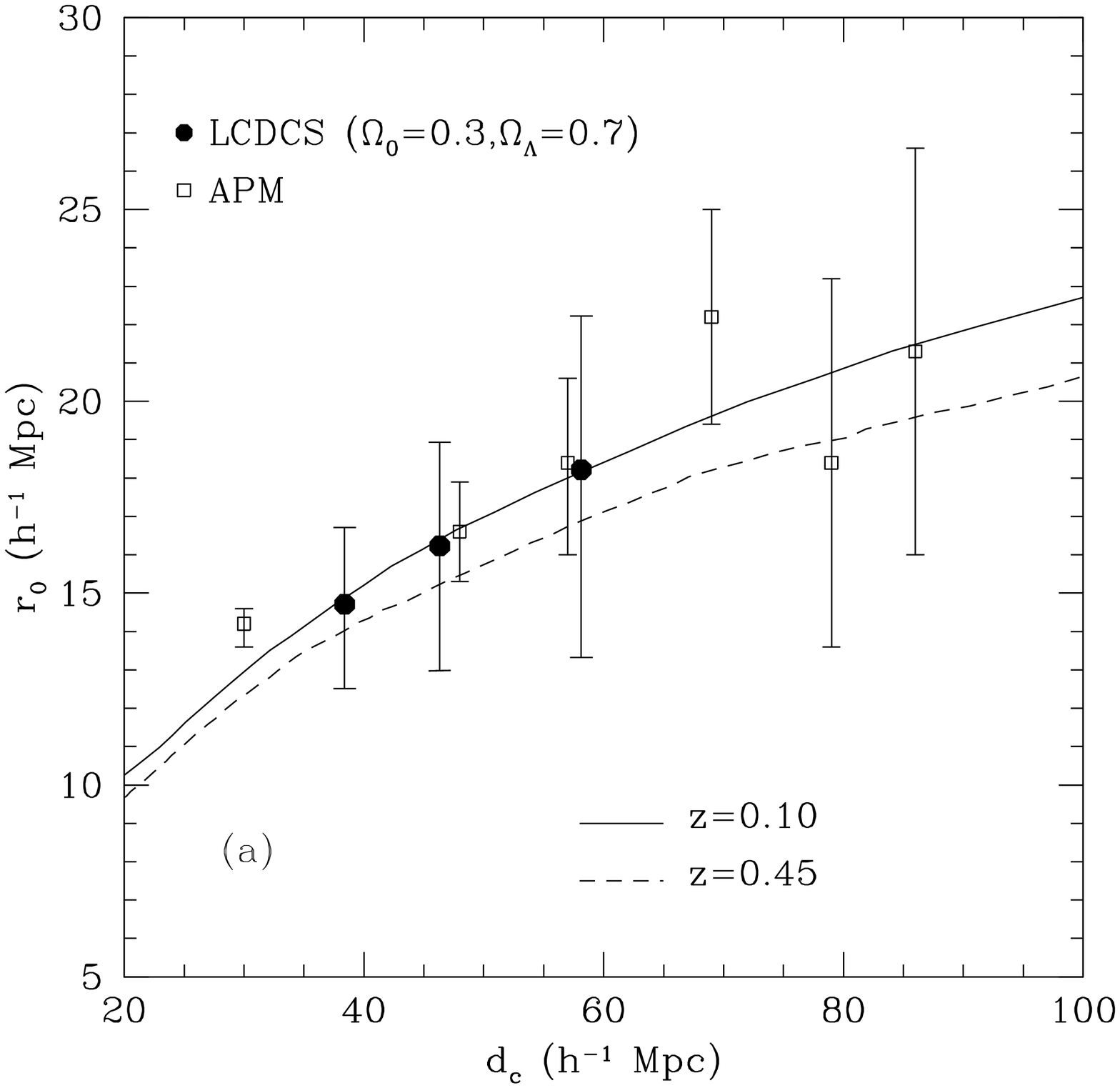} 
\plotone{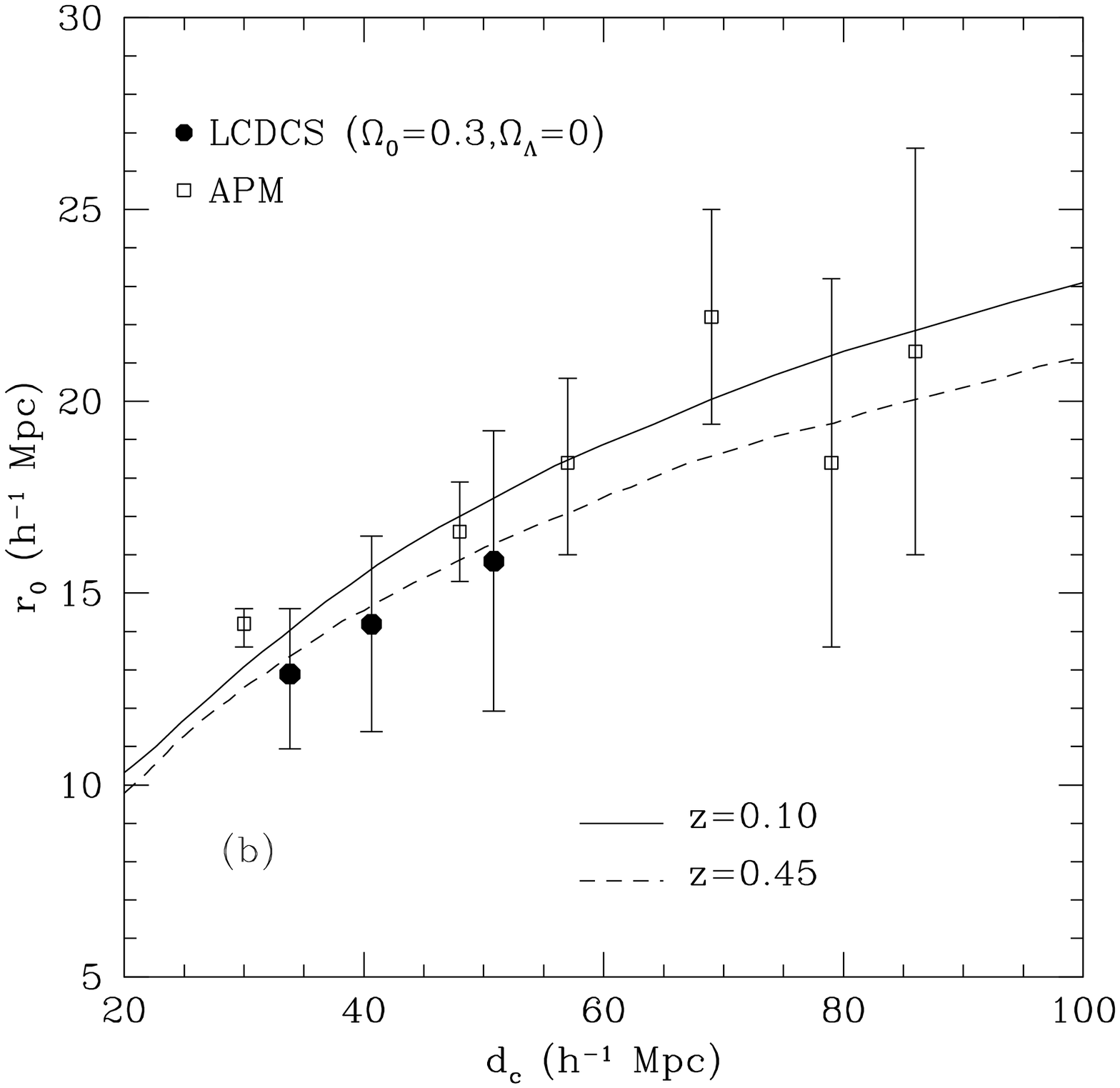} 
\plotone{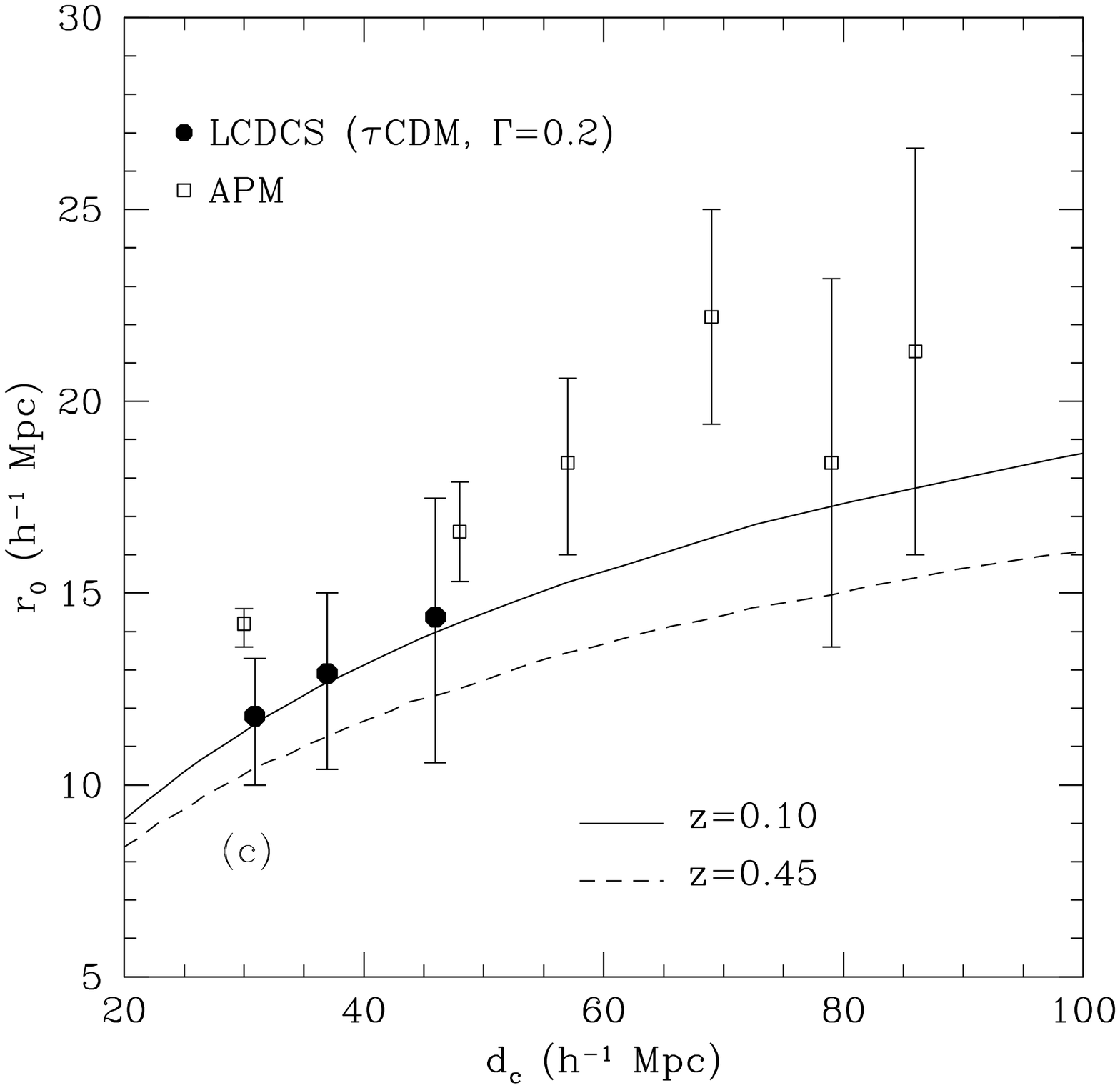}
\figcaption[apmcomp] {Comparison of the LCDCS and APM data sets
with analytic predictions. Different panels correspond to the
different cosmologies assumed in deriving the LCDCS correlation 
lengths (see Table \ref{tab:rodc}). The analytic models include
the impact of uncertainty in clusters masses for the two data sets,
which causes these curves to be slightly shallower than the fiducial
models in Figure \ref{fig:cfvslocal}. 
\label{fig:apmcomp}}
\end{figure*}            

It is evident in Figure \ref{fig:cfvslocal} that the weak evolution in the
correlation length predicted
for this $\Lambda$CDM model
is consistent with the observations; however, for
a more accurate comparison we should incorporate into the models the impact
of observational scatter in the subsample mass thresholds for the different
data sets. We restrict our attention to the APM and
LCDCS catalogs in this analysis. 
For the APM catalog, we use the correlation lengths from \citet{cro97}, taking
the velocity dispersion scatter about the sample thresholds 
to be d$\log \sigma$=0.13 \citep[based upon the data set of][]{alo99}. We estimate 
the mean redshift of the APM samples to be $<$$z$$>$$\simeq$0.1.
For the LCDCS catalog we use the
surface brightness-velocity dispersion relation and surface brightness scatter
from \citet{gon2001}. The data and resulting models that include 
this
scatter are shown in Figure \ref{fig:apmcomp}$a$-$c$ for the three cosmologies
discussed in this paper. Both the APM and LCDCS data are consistent with the
low-density models. For the plotted curves,  the combined
data sets yield reduced $\chi^2$ values $\chi_\nu^2$=0.7 and $\chi_\nu^2$=0.6 for the $\Lambda$CDM and OCDM
models, respectively.\footnote{The bulk of the weight in computing
these $\chi_\nu^2$ values comes from the APM data set. The LCDCS provides only
slight additional leverage due to the large statistical uncertainties associated
with these measurements.} 
 In contrast, the plotted $\tau$CDM model systematically
underpredicted the $r_0$ values for both data sets ($\chi^2$=4.1 for $\Gamma$=0.2). 
$\tau$CDM can be made to better match the data by decreasing $\Gamma$;
however, it is not possible to make $\tau$CDM simultaneously
consistent with the cluster correlation length data and the galaxy
power spectrum (e.g. \citealt{eis2000}).

\section{Systematic Uncertainties}

It is important to consider whether any systematic biases have the
potential to qualitatively alter our results.  Table
\ref{tab:systematics} summarizes the systematic effects discussed thus far
in the text, as well as several additional potentially important factors.
The most significant potential systematic bias is attributable to the
impact of large scale structure. The LCDCS serendipitously includes
the most X-ray luminous cluster known 
\citep[RX J1347.5$-$1145 at $z$=0.45;][]{sch95}.
Three of the ten
most massive LCDCS candidates with estimated redshifts $z_{est}$$\le$0.58 have
projected separations of less than 50 $h^{-1}$ Mpc relative to RX
J1347.5$-$1145, indicating that this region likely contains a massive
supercluster.
To estimate the sensitivity of our results to
the presence of this region within the survey, we recompute the 
correlation lengths excluding candidates within 100$h^{-1}$ Mpc of 
RX J1347.5$-$1145. For all three subsamples the results change by less than 15\%
(i.e. less than $1-\sigma$), and for the lowest redshift subsample the correlation length actually
increases slightly when this region is excluded. We thus conclude it likely that
no individual supercluster should systematically alter our results by more
that 15\%.

\begin{inlinetable} 
\begin{center}
\caption{Systematic uncertainties} \label{tab:systematics}
\begin{tabular}{lr}
\hline\hline
\multicolumn{1}{c}{Issue} &  \multicolumn{1}{c}{Impact} \\
\hline
\multicolumn{2}{r}{\underline{$r_0$}}\\
\\
Large scale structure & $\la$15\% \\
Fractional contamination & 6\% \\
Clustering strength of contamination & $\la$10\% \\
Model for $dN/dz$ in Limber equation & 3\% \\
Extinction-dependent selection effects & 1\% \\
Redshift dependence of $\Sigma$ & $<$1\% \\
Fixing $\gamma$. Would increase $r_0$ for subsamples & 0-10\% \\
$\;\;\;$ (larger change for higher $d_c$ bins)&\\
\\
\multicolumn{2}{r}{\underline{$d_c$}}\\
\\
Uncertainty in effective comoving volume & $<$3\%\\
\hline
\end{tabular}
\end{center}
\begin{minipage}{0.60\linewidth}
\end{minipage}
\end{inlinetable}

 A second key concern, due to the cluster detection method, is whether
there exist extinction-dependent selection effects that impact the
derived correlation lengths. As a test we recompute $r_0$ for the lowest 
redshift subsample using only regions with E($\bv$)$\le$0.06, which
reduces the sample from 178 to 89 clusters. The recovered correlation length for
these low-extinction regions is 13.2 (vs. 14.7 for all 178 clusters). This 10\%
change is less than the statistical uncertainty,
indicating that our results are not strongly sensitive to variations in the
galactic extinction within the survey region. 

Of the factors previously discussed in the text, the greatest potential
impact upon our results can be achieved if the {\it a priori} assumption
is made that all subsamples have the same power law slope, $\gamma$.
By making this assumption we test whether our results are robust to any
factor that may be artificially enhancing the best-fit values of $\gamma$.
 If we fix $\gamma$=2.15, 
then the derived $r_0$ values for three subsamples increase by 0\%, 6\%,
and 15\% (in order of increasing maximum redshift for the subsamples). While
these changes are less than the statistical uncertainties, they do act to 
steepen the slope of the observed $r_0$-$d_c$ relation and thus slightly
degrade the agreement with low-density models.

We next consider two additional issues that might qualitatively change
the results of this paper. First, what if the redshift uncertainty is
significantly larger than indicated by \citet{gon2001}? In this case
the redshift ranges spanned by our 
subsamples are larger --- and more similar to one another --- than our
current best assessment.  If true, then we are currently
underestimating $r_0$. Quantitatively, if $\sigma_z/z$=0.25 instead of
0.14, then the actual values of $r_0$ would be $\sim$2 $h^{-1}$ Mpc
larger.  On the other hand, neither the mean separation $d_c$ for a
given sample or the steepness of the relation between $r_0$ and $d_c$
changes perceptibly. Second, what if the contamination fraction is
uniform in all the subsamples rather than being an increasing function
of redshift? Qualitatively, this would imply that part of the observed
weak dependence of $r_0$ upon $d_c$ is simply due to our contamination
correction. Quantitatively, if we take the mean contamination rate to
be 16\% for all three subsamples, then the $r_0$ values($\Lambda$CDM) 
change from
(14.7,16.2,18.2) to (15.0,16.2,17.7) $h^{-1}$ Mpc -- a minor effect
compared with other uncertainties.

From our analysis of the above systematics, we conclude that the two
largest systematic uncertainties in this work arise from large scale
structure and potential underestimation of the redshift uncertainty.
Large scale structure can impact the derived values of $r_0$ at the
 $1-\sigma$, while underestimation of the redshift uncertainty could 
yield correlation lengths that are systematically too small (by up to $\sim$2
$h^{-1}$ Mpc if $\sigma_z/z$=0.25). 

\section{Discussion and Conclusions}
\label{sec-cosmodiscussion}

The Las Campanas Distant Cluster Survey is the largest existing
catalog of clusters at $z$$>$0.3, providing a unique sample with which
to study the properties of the cluster population.  We have used the
LCDCS to constrain the cluster-cluster angular correlation function,
providing the first measurements for a sample with a mean redshift
$z$$\ga$0.2.  From the observed angular correlation function, we
derive the spatial correlation length, $r_0$, as a function of mean
separation, $d_c$.  
We find that the LCDCS correlation
lengths are in agreement with results from local samples, and observe
a dependence of $r_0$ upon $d_c$ that is comparable to the results of
\citet{cro97} for the APM catalog.  This clustering strength, its
dependence on number density, and its minimal redshift evolution are
consistent with analytic expectations for low density models, and with
results from the $\Lambda$CDM Hubble Volume simulations.
Consequently, while statistical uncertainty limits our ability to
discriminate between cosmological models, our results are in
concordance with the flat $\Lambda$CDM model favored by recent
supernovae and cosmic microwave background observations
\citep[e.g.,][]{riess01,pryke01,boom01}.

\section{Acknowledgments}
The authors thank the anonymous referee for a thorough report that
significantly improved this paper.
AHG acknowledges support from the NSF Graduate Research Fellowship
Program, the ARCS Foundation, and the Harvard-Smithsonian Center for
Astrophysics.  DZ acknowledges financial support from NSF CAREER grant
AST-9733111, and fellowships from the David and Lucile Packard
Foundation and Alfred P. Sloan Foundation.  RHW was supported by a
GAANN fellowship at UCSC.

\appendix
\section{Analytic Models for the Spatial Correlation Function}

Analytic models for the spatial correlation function are based upon
the Press-Schechter formalism \citep[PS;][]{pre74} formalism.
\citet{mo96} were the first to use this formalism to derive the
expected correlation for mass-dependent bias, finding that if 
the observed correlation function is defined as
\begin{equation}
\xi(r,z,M)=b^2(z,M) \xi_{m}(r,z),
\end{equation}
where $\xi_{m}$ is the average mass correlation function,
\begin{equation}
\xi_{m}(r,z)=D_+^2(z)\int_0^\infty P(k) \frac{\sin kr}{kr} d^3 k,
\end{equation}
then the bias, $b(z,M)$, is given by the equation
\begin{equation}
b(z,M)=1+\frac{\delta_c(z)}{\sigma^2(M)} - \frac{1}{\delta_c(z)}  .
\label{eqn:bias}
\end{equation}
  Modeling of observed correlation functions requires calculation of
the effective bias, $b_{eff}(z,M)$, which is the average of $b(z,M)$
weighted by the mass function of the clusters used to compute the
correlation function \citep{mo96,mat97,bau98},
\begin{equation}
b_{eff}=\frac{\int_m^\infty b(M)n(M)dM}{\int_m^\infty n(M)dM},
\end{equation}
where $m$ is the mass limit of the cluster sample.

This prescription can also be easily extended to incorporate the
\citeauthor{she99} modifications to PS by replacing equation
(\ref{eqn:bias}) with
\begin{equation}
b(z,M)=1+\frac{1}{\delta_c(0)}\left[\frac{a\delta^2_c(z)}{\sigma^2}-1\right] +
\frac{2p}{\delta_c(0)}\left[1+\left(\frac{\sqrt{a}\delta_c}{\sigma}\right)^{2p}\right]^{-1}
\end{equation} 
where $a$=0.707 and $p$=0.3 \citep{mos2000}.  This is the analytic
model used in \S\ref{subsec-cfres}. For comparison, we also overlay
results from the Virgo Consortium Hubble Volume simulations
\citep{col2000} in Figure \ref{fig:cfvslocal}$a$.

\end{document}